\documentclass[onecolumn,showpacs]{revtex4}

\usepackage{graphicx}
\usepackage{dcolumn}
\usepackage{bm}

\input epsf

\begin{document}

\title{\Large Thermodynamics in Quasi-Spherical Szekeres Space-Time}

\author{\bf Ujjal Debnath\footnote{ujjaldebnath@yahoo.com ,
ujjal@iucaa.ernet.in}}

\affiliation{Department of Mathematics, Bengal Engineering and
Science University, Shibpur, Howrah-711 103, India.\\}

\date{\today}

\begin{abstract}
We have considered that the universe is the inhomogeneous $(n+2)$
dimensional quasi-spherical Szekeres space-time model. We consider
the universe as a thermodynamical system with the horizon surface
as a boundary of the system. To study the generalized second law
(GSL) of thermodynamics through the universe, we have assumed the
trapped surface is the apparent horizon. Next we have examined the
validity of the generalized second law of thermodynamics (GSL) on
the apparent horizon by two approaches: (i) using first law of
thermodynamics on the apparent horizon and (ii) without using the
first law. In the first approach, the horizon entropy have been
calculated by the first law. In the second approach, first we have
calculated the surface gravity and temperature on the apparent
horizon and then horizon entropy have found from area formula. The
variation of internal entropy have been found by Gibb's law. Using
these two approaches separately, we find the conditions for
validity of GSL in $(n+2)$ dimensional quasi-spherical Szekeres
model.
\end{abstract}

\pacs{}

\maketitle

\section{\normalsize\bf{Introduction}}

In Einstein gravity, the evidence of the connection between
gravity and thermodynamics was first discovered in [1] by deriving
the Einstein equation from the proportionality of entropy and
horizon area together with the first law of thermodynamics. The
horizon area of black hole is associated with its entropy, the
surface gravity is related with its temperature in black hole
thermodynamics [2]. Then Padmanabhan [3] was able to formulate the
first law of thermodynamics on the horizon, starting from Einstein
equations for a general static spherically symmetric space time.
Frolov et al [4] calculated the energy flux of a background
slow-roll scalar field through the quasi-de Sitter apparent
horizon and used the first law of thermodynamics $-dE = TdS$,
where $dE$ is the amount of the energy flow through the apparent
horizon. Using the Hawking temperature $T_{A}=\frac{1}{2\pi
R_{A}}$ and Bekenstein entropy $S_{A}=\frac{\pi R_{A}^{2}}{G}$
($R_{A}$ is the radius of apparent horizon) at the apparent
horizon, the first law of thermodynamics (on the apparent horizon)
is shown to be equivalent to Friedmann equations [5] and the
generalized second law of thermodynamics is obeyed at the horizon.
The thermodynamics in de Sitter space–time was first investigated
by Gibbons and Hawking in [6]. When the apparent horizon and the
event horizon of the Universe are different, it was found that the
first law and generalized second law (GSL) of thermodynamics hold
on the apparent horizon, while they break down if one considers
the event horizon [7]. On the basis of the well known
correspondence between the Friedmann equation and the first law of
thermodynamics of the apparent horizon, Gong et al [8] argued that
the apparent horizon is the physical horizon in dealing with
thermodynamics problems. Considering FRW model of the universe,
most studies deal with validity of the generalized second law of
thermodynamics starting from the first law when universe is
bounded by the apparent horizon [9]. But there are few works of
the justification of first and second laws of thermodynamics on
the event horizon [10]. The validity of thermodynamical laws in
generalized gravity theories have also discussed in [11].\\

Usually, for cosmological phenomena over galactic scale or in the
smaller scale, it is reasonable to consider inhomogeneous
solutions to Einstein equations. Szekeres' [12] in 1975, gave a
class of inhomogeneous solutions representing irrotational dust.
The space-time represented by these solutions has no killing
vectors and it has invariant family of spherical hypersurfaces.
Hence this space-time is referred as quasi-spherical space-time.
Subsequently, the solutions have been extended by Szafron [13] and
Szafron and Wainwright [14] for perfect fluid and they studied
asymptotic behaviour for different choice of the parameters
involved. Later Barrow and Stein-Schabes [15] gave solutions for
dust model with a cosmological constant and showed the validity of
the Cosmic `no-hair' Conjecture. Recently, Chakraborty et al [16]
have extended the Szekeres solution to $(n+2)$ dimensional
space-time and generalized it for matter containing heat flux
[17]. Recently, several works have been done on gravitational
collapse using this higher dimensional Szekeres solution [18,19]. \\

In this work, we consider the $(n+2)$ dimensional quasi-spherical
Szekeres space-time. Next we'll examine the validity of the
generalized second law of thermodynamics (GSL) on the apparent
horizon by two approaches: (i) using first law of thermodynamics
on the apparent horizon and (ii) without using first law. In the
first approach, we don't need the horizon temperature. So the
horizon entropy can be calculated from the first law. In the
second approach, first we calculate surface gravity and
temperature on the apparent horizon and then horizon entropy can
be found from area formula. Using these two approaches, we find
the conditions for validity of GSL in quasi-spherical Szekeres model.\\

\section{\normalsize\bf{The Szekeres' Model}}

The metric ansatz for the $(n+2)$ dimensional Szekeres' space-time
[12, 16] is of the form

\begin{equation}
ds^{2}=-dt^{2}+e^{2\alpha}dr^{2}+e^{2\beta}\sum_{i=1}^{n}dx_{i}^{2}
\end{equation}

where the metric coefficients $\alpha$ and $\beta$ are functions
of all space-time co-ordinates i.e.,
$$\alpha=\alpha(t,r,x_{1},....,x_{n}),~~ \beta=\beta(t,r,x_{1},....,x_{n}).$$

Now Considering both radial and transverse stresses the energy
momentum tensor has the structure

$$T_{\mu}^{\nu}=\text{diag}(-\rho,p_{r},p_{_{T}},...,p_{_{T}})$$

Now for the choice namely $\beta'\ne 0,~ \dot{\beta}_{x_{i}}=0$ we
have from the field equations the explicit form of the metric
coefficients are as follows [16]:

\begin{equation}
e^{\beta}=R(t,r)~e^{\nu(r,x_{1},...,x_{n})}
\end{equation}

and

\begin{equation}
e^{\alpha}=\frac{R'+R~\nu'}{\sqrt{1+f(r)}}
\end{equation}

and the evolution equation for $R$ gives

\begin{equation}
R\ddot{R}+\frac{1}{2}{(n-1)\dot{R}^{2}}+\frac{p_{r}}{n}{R^{2}}=\frac{n-1}{2}~f(r),
\end{equation}

where $f(r)$ is the function of $r$. Also the function $\nu$
satisfies

 \begin{equation}
 e^{-2\nu}\sum_{i=1}^{n}[{(n-2)\nu_{x_{i}}^{2}+2\nu_{x_{i}x_{i}}}]=-n
 \end{equation}
 which has a solution of the form
 \begin{equation}
e^{-\nu}=A(r)\sum_{i=1}^{n}x_{i}^{2}+\sum_{i=1}^{n}B_{i}(r)x_{i}+C(r)
 \end{equation}
with the restriction,
 \begin{equation}
\sum_{i=1}^{n}B_{i}^{2}-4AC=-1
\end{equation}

for the arbitrary functions $A(r)$, $B_{i}(r),(i=1,2,..,n)$ and
$C(r)$.\\

Now from conservation equation $T^{\nu}_{\mu ;~\nu}=0$ we get [19]

\begin{equation}
\left.
\begin{array}{c}
\dot{\rho}+\dot{\alpha}(\rho+p_{r})+n\dot{\beta}(\rho+p_{_{T}})=0\\\\
p_{r}'+n\beta'(p_{r}-p_{_{T}})=0\\\\
\text{and}~~~ \alpha_{x_{i}}(p_{r}-p_{_{T}})=
\frac{\partial}{\partial x_{i}
}p_{_{T}}~~~~(i=1,2,...,n)\\\\
\end{array}
\right\}
\end{equation}

In the general case when both radial and tangential pressures are
non-zero and distinct then from the Einstein equations
$G_{\mu\nu}=T_{\mu\nu}$ (choosing $8\pi G=c=1$) they can be
obtained in compact form as [19]

\begin{equation}
\rho=\frac{F'}{\zeta^{n}\zeta'}~~,~~p_{r}=-\frac{\dot{F}}{\zeta^{n}\dot{\zeta}}~~\text{and}~~
p_{_{T}}=p_{r}+\frac{\zeta p_{r}'}{n\zeta'}
\end{equation}

where

\begin{equation}
\zeta=e^{\beta}   ~~\text{and}~~
F(t,r)=\frac{n}{2}R^{n-1}e^{(n+1)\nu}(\dot{R}^{2}-f(r))
\end{equation}

Now consider radial and tangential pressures are equal i.e.,
$p_{r}=p_{_{T}}=p$ , so from (8), we get the isotropic pressure is
function of $t$ only i.e., $p=p(t)$. As there is no restriction on
the energy density so $\rho$ is in general a function of all the
($n+2$) variables i.e., $\rho=\rho(t,r,x_{1},...,x_{n})$ and hence
no equation of state is imposed. So the conservation equation (8)
yields to

\begin{equation}
\dot{\rho}+(\dot{\alpha}+n\dot{\beta})(\rho+p)=0
\end{equation}

Also the metric (1) can be written as

\begin{equation}
ds^{2}=-dt^{2}+\frac{(R'+R\nu')^{2}}{1+f(r)}dr^{2}+R^{2}e^{2\nu}\sum_{i=1}^{n}dx_{i}^{2}
\end{equation}

Here $R$ is the radius of the non-concentric spheres. If the
spheres are concentric i.e, if $\nu'=0$ then the above Szekeres
metric reduces to $(n+2)$ dimensional spherically symmetric
Lama$\hat{\i}$tre-Tolman-Bondi (LTB) metric [16]. \\

\section{\normalsize\bf{GSL of Thermodynamics on the Apparent Horizon}}

Now we consider the metric (8) in the following form

\begin{equation}
ds^{2}
=h_{ab}dx^{a}dx^{b}+R^{2}e^{2\nu}\sum_{i=1}^{n}dx_{i}^{2}~~,~~a,b=0,1
\end{equation}

where $h_{ab}=diag\left(-1,\frac{(R'+R\nu')^{2}}{1+f(r)}\right)$.\\

The formation of event horizon depends greatly on the computation
of null geodesics whose computation are almost impracticable for
the present space-time geometry. So a closely related concept of a
trapped surface (a space-like 2-surface whose normals on both
sides are future pointing converging null geodesic families) will
be considered. The dynamical apparent horizon $R_{A}$, a
marginally trapped surface with vanishing expansion, is determined
by the relation [5,19,21] (see also {\bf APPENDIX})

\begin{equation}
h^{ab}\partial_{a}(Re^{\nu})\partial_{b}(Re^{\nu})=0
\end{equation}

This implies

\begin{equation}
\dot{R}_{A}^{2}=1+f(r)
\end{equation}

So from equation (10), we obtain (on the apparent horizon)

\begin{equation}
F(t,r)=\frac{n}{2}R_{A}^{n-1}e^{(n+1)\nu}
\end{equation}

Now the Gibb's law of thermodynamics states that [7]

\begin{equation}
T_{A}dS_{I}=pdV+d(E_{I})
\end{equation}

where, $S_{I},~p,~V$ and $E_{I}$ are respectively entropy,
pressure, volume and internal energy within the apparent horizon.
Here the expression for internal energy can be written as
$E_{I}=\rho V$. Here $T_{A}$ is the temperature on the apparent
horizon. Now the volume of the $(n+1)$ dimensional space is [5]

\begin{equation}
V=\Omega_{n+1} R_{A}^{n+1}e^{(n+1)\nu}~~ \text{where} ~~
\Omega_{n+1}=\frac{\pi^{\frac{n+1}{2}}}{\Gamma(\frac{n+3}{2})}
\end{equation}

The time variation of internal entropy is obtained as (using (2),
(3), (11), (17) and (18))

\begin{equation}
\dot{S}_{I}=\frac{\Omega_{n+1} R_{A}^{n+1}e^{(n+1)\nu}}{T_{A}}~
(\rho+p)
\left(\frac{\dot{R}_{A}}{R_{A}}-\frac{\dot{R}'_{A}+\dot{R}_{A}\nu'}{R'_{A}+R_{A}\nu'}
\right)
\end{equation}

\subsection{\large\bf{Validity Conditions of GSL using First Law of Thermodynamics}}

The unified first law is defined by [22]

\begin{equation}
dE={\cal A}\Psi+WdV
\end{equation}

where

\begin{equation}
{\cal A}=(n+1)\Omega_{n+1} R^{n}e^{n\nu}
\end{equation}

is the area [5] and the volume $V$ is defined in (18). The work
density function is given by

\begin{equation}
W=-\frac{1}{2}h^{ab}T_{ab}=\frac{1}{2}(\rho-p)
\end{equation}

The energy-supply vector is given by

\begin{equation}
\Psi_{a}=h^{bc}T_{ac}\partial_{b} (Re^{\nu})+W\partial_{a}
(Re^{\nu})=\left(-\frac{1}{2}(\rho+p)\dot{R}e^{\nu},
\frac{1}{2}(\rho+p)(R'+R\nu')e^{\nu} \right)
\end{equation}

So

\begin{equation}
\Psi=\Psi_{a}dx^{a}=-\frac{1}{2}(\rho+p)e^{\nu}[\dot{R}dt-(R'+R\nu')dr]
\end{equation}

The total energy inside the quasi-spherical surface is given by

\begin{equation}
E=\frac{n(n+1)}{2}\Omega_{n+1}
R^{n-1}e^{(n-1)\nu}[e^{2\nu}-h^{ab}\partial_{a}(Re^{\nu})\partial_{b}(Re^{\nu})]=\frac{n(n+1)}{2}\Omega_{n+1}
R^{n-1}e^{(n+1)\nu}[\dot{R}^{2}-f(r)]
\end{equation}

Comparing (10) and (25), we get
\begin{equation}
E=(n+1)\Omega_{n+1}F
\end{equation}

From this, we can say that $F(t,r)$ represents the mass function
within the quasi-spherical surface. Now using (18), (21), (22) and
(24), we get

\begin{equation}
A\Psi+WdV=(n+1)\Omega_{n+1}R^{n}e^{(n+1)\nu}[-p\dot{R}dt+\rho
(R'+R\nu')dr]
\end{equation}

Using (25) and (27), comparing the coefficients of $dt$ and $dr$
in (20), we can recover the field equations (4) and (9). Now from
the unified first law (20) and using (27), we get

\begin{equation}
dE=(n+1)\Omega_{n+1}R^{n}e^{(n+1)\nu}[-(\rho+p)\dot{R}dt+\rho
e^{-\nu}d(Re^{\nu})]
\end{equation}

We know that heat is one of the forms of energy. Therefore, the
heat flow $\delta Q$ through the apparent horizon is just the
amount of energy crossing it during the time interval $dt$. That
is, $\delta Q=-dE$ is the change of the energy inside the apparent
horizon. So the amount of the energy crossing on the apparent
horizon is given by [20]

\begin{equation}
-dE_{A}=(n+1)\Omega_{n+1}R_{A}^{n}\dot{R}_{A}e^{(n+1)\nu}(\rho+p)dt={\cal
A}\dot{R}_{A}e^{\nu}T_{\mu\nu}k^{\mu}k^{\nu}dt
\end{equation}

The first law of thermodynamics (Clausius relation) on the
apparent horizon is defined as follows:

\begin{equation}
T_{A}dS_{A}=dQ=-dE_{A}
\end{equation}

So using (29) and (30), we obtain the time variation of the
entropy on the apparent horizon as

\begin{equation}
\dot{S}_{A}=\frac{(n+1)\Omega_{n+1}}{T_{A}}~
R_{A}^{n}\dot{R}_{A}e^{(n+1)\nu}(\rho+p)
\end{equation}

Combining (19) and (31), we obtain

\begin{equation}
\dot{S}_{I}+\dot{S}_{A}=\frac{\Omega_{n+1}
R_{A}^{n+1}e^{(n+1)\nu}}{T_{A}}~ (\rho+p)
\left((n+2)\frac{\dot{R}_{A}}{R_{A}}-\frac{\dot{R}'_{A}+\dot{R}_{A}\nu'}{R'_{A}+R_{A}\nu'}
\right)
\end{equation}

Using (2), (3), (9), (15), (16) and (32), after manipulation we
get the rate of change of total entropy as

\begin{eqnarray*}
\dot{S}_{I}+\dot{S}_{A}=\frac{\Omega_{n+1} F}{T_{A}} \left[
\left(\left(\frac{n}{2}\right)^{\frac{1}{n-1}}
\frac{(n-1)F'\sqrt{1+f}}{F'-2F\nu'}F^{\frac{n-2}{n-1}}e^{\frac{n+1}{n-1}\nu}-\dot{F}
\right) \times \right.
\end{eqnarray*}

\begin{equation}
\left. ~~~~~~~~~~~~~~~~~~~~~~~~~~~~~
\left(\frac{(n+2)}{F}-\frac{(n-1)(f'+2(1+f)\nu')}{2(1+f)~(F'-2F\nu')}
\right) \right]
\end{equation}

If the expression inside the square bracket is non-negative then
the GSL will be justified. For marginally bound case, i.e., for
$f(r)=0$, the GSL is satisfied if the following conditions hold:\\

~~~~~~~~~~~~~~~~~~~~~~~~~~~~(i)  $F'\ge \frac{3(n+1)}{n+2}F\nu'$
and $\dot{F}\le
3(n+1)\left(\frac{n}{2}\right)^{\frac{1}{n-1}}F^{\frac{n-2}{n-1}}e^{\frac{n+1}{n-1}\nu}$\\

~~~~~~~~~~~~~~~~~OR,~~~~ (ii) $F'\le \frac{3(n+1)}{n+2}F\nu'$ and
$\dot{F}\ge
3(n+1)\left(\frac{n}{2}\right)^{\frac{1}{n-1}}F^{\frac{n-2}{n-1}}e^{\frac{n+1}{n-1}\nu}$.

\subsection{\large\bf{Validity Conditions of GSL without using First Law of Thermodynamics}}

The surface gravity is defined as

\begin{equation}
\kappa=\frac{1}{2\sqrt{-h}}\partial_{a}(\sqrt{-h}h^{ab}\partial_{b}(Re^{\nu}))
\end{equation}

where $h={\det}(h_{ab})$. So on the apparent horizon, we get

\begin{equation}
\kappa=-\frac{\ddot{R}_{A}e^{\nu}}{2}-\frac{\dot{R}_{A}(\dot{R}'_{A}+\dot{R}_{A}\nu')e^{\nu}}{2(R'+R\nu')}
+\frac{\sqrt{1+f}}{2(R'+R\nu')}\frac{\partial}{\partial
r}(e^{\nu}\sqrt{1+f})
\end{equation}

Now apparent horizon temperature is (using (3)-(5), (9), (10),
(15) and (16))

\begin{equation}
T_{A}=\frac{|\kappa|}{2\pi}=\frac{e^{\nu}}{8\pi}\left|
(n-1)\left(\frac{n}{2F}\right)^{\frac{1}{n-1}}e^{\frac{n+1}{n-1}\nu}-
\frac{\dot{F}}{F\sqrt{1+f}}\right|
\end{equation}

Since one can relate the entropy with the surface area of the
apparent horizon through $S_{A} = {\cal A}/4G$. Therefore using
(21) we have

\begin{equation}
S_{A}=2\pi(n+1)\Omega_{n+1}
R_{A}^{n}e^{n\nu}~~~,~~~~~~~~~~(\text{since}~~8\pi G=1)
\end{equation}

So the variation of entropy on the apparent horizon is obtained as
(using (15) and (16))

\begin{equation}
\dot{S}_{A}=4\pi(n+1)\Omega_{n+1}F\sqrt{1+f}~e^{-\nu}
\end{equation}

Using (9), (10), (15), (16), (19), (36) and (38), we finally
obtain (after manipulation) the rate of change of total entropy as

\begin{eqnarray*}
\dot{S}_{I}+\dot{S}_{A}=\frac{\Omega_{n+1} F}{T_{A}} \left[
\left(\left(\frac{n}{2}\right)^{\frac{1}{n-1}}
\frac{(n-1)F'\sqrt{1+f}}{F'-2F\nu'}F^{\frac{n-2}{n-1}}e^{\frac{n+1}{n-1}\nu}-\dot{F}
\right) \times \right.
\end{eqnarray*}

\begin{equation}
\left.
\left(\frac{1}{F}-\frac{(n-1)(f'+2(1+f)\nu')}{2(1+f)~(F'-2F\nu')}
\right)+\frac{(n+1)\sqrt{1+f}}{2}\times \left|
(n-1)\left(\frac{n}{2F}\right)^{\frac{1}{n-1}}e^{\frac{n+1}{n-1}\nu}-
\frac{\dot{F}}{F\sqrt{1+f}}\right|~~ \right]
\end{equation}

If the expression inside the square bracket is non-negative then
the GSL will be justified. From the above expression, we cannot
draw any definite conclusion.\\

\section{\normalsize\bf{Discussions}}

We have considered that the universe is the inhomogeneous $(n+2)$
dimensional quasi-spherical Szekeres space-time model. We consider
the universe as a thermodynamical system with the horizon surface
as a boundary of the system. To study the generalized second law
(GSL) of thermodynamics through the universe, we have assumed the
trapped surface is the apparent horizon. Next we have examined the
validity of the generalized second law of thermodynamics (GSL) on
the apparent horizon by two approaches: (i) using first law of
thermodynamics on the apparent horizon and (ii) without using the
first law. In the first approach, the horizon entropy have been
calculated by the first law. In the second approach, first we have
calculated the surface gravity and temperature on the apparent
horizon and then horizon entropy have found from area formula. The
variation of internal entropy have been found by Gibb's law. Using
these two approaches separately, we find the conditions for
validity of GSL in $(n+2)$ dimensional quasi-spherical Szekeres
model. Also  for marginally bound case, we have found the bounds
on the derivatives of the mass function $F$.\\

$${\bf APPENDIX}$$

Define, $X=\frac{R'+R\nu'}{\sqrt{1+f(r)}}$ and $Y=Re^{\nu}$. Let
us consider the 2-surface $S_{r,t}$ ($r=$constant, $t=$constant)
is a trapped surface and $K^{\mu}$ denotes the tangent vector
field to the null geodesics which is normal to $S_{r,t}$. So on
the apparent horizon (on $S_{r,t}$) we have [21]

\begin{equation}
K_{\mu}~K^{\mu}=0,~~K^{\mu}_{~;~\nu}~K^{\nu}=0
\end{equation}

and

\begin{equation}
K^{2}=K^{3}=0,~~(K^{0})^{2}-X^{2}(K^{1})^{2}=0
\end{equation}

Now on $S_{r,t}$, the choice of affine parameter may clearly be
such that

\begin{equation}
K^{0}=X,~K^{1}=\epsilon=\pm 1
\end{equation}

Now on $S_{r,t}$,

\begin{equation}
K^{\mu}_{;\mu}=K^{\mu}_{,\mu}+\Gamma^{\mu}_{\mu\nu}K^{\nu}
=K^{0}_{,0}+K^{1}_{,1}+X\left(\frac{\dot{X}}{X}+\frac{\dot{Y}}{Y}\right)
+\epsilon \left(\frac{X'}{X}+\frac{Y'}{Y} \right)
\end{equation}

Since $K^{2}_{,2}=K^{3}_{,3}=0$ on $S_{r,t}$. On the other hand,
forming $\partial/\partial t$ of the first equation of (32) and
setting $\mu=1$ in the second equation gives on $S_{r,t}$

\begin{equation}
K^{0}_{,0}-\epsilon X K^{1}_{,0}-\dot{X}=0
\end{equation}
and
\begin{equation}
K^{1}_{,0}X+\epsilon (K^{1}_{,1}+2\dot{X})+\frac{X'}{X}=0
\end{equation}

Eliminating $K^{1}_{0}$ between these two equations and
substituting in (35) gives

\begin{equation}
K^{\mu}_{;\mu}=\frac{2}{Y}(X\dot{Y}+\epsilon
Y')=\frac{2(R'+R\nu')}{R}\left(\frac{\dot{R}}{\sqrt{1+f(r)}}+\epsilon\right)
\end{equation}

On apparent horizon, $K^{\mu}_{;\mu}=0$ gives

\begin{equation}
~~~~~~~~~~~~~~~~~~~\frac{\dot{R}}{\sqrt{1+f(r)}}+\epsilon=0
~~~~(\text{since}, ~~R'+R\nu'\ne 0)
\end{equation}
\begin{equation}
\Rightarrow ~~~\dot{R}^{2}=\epsilon^{2}(1+f(r))=1+f(r)
\end{equation}\\

{\bf References:}\\
\\
$[1]$ T. Jacobson, {\it Phys. Rev. Lett.} {\bf 75} 1260 (1995).\\
$[2]$ J. D. Bekenstein, {\it Phys. Rev. D} {\bf 7} 2333 (1973); S.
W. Hawking, {\it Commun. Math. Phys.} {\bf 43} 199 (1975); J. M.
Bardeen, B. Carter and S. W. Hawking, {\it Commun. Math. Phys.}
{\bf 31} 161 (1973).\\
$[3]$ T. Padmanabhan, {\it Class. Quantum Grav.} {\bf 19} 5387
(2002).\\
$[4]$ A. V. Frolov and L. Kofman, {\it JCAP} {\bf 0305} 009
(2003).\\
$[5]$ R. G. Cai and S. P. Kim, {\it JHEP} {\bf 02} 050 (2005).\\
$[6]$ G. W. Gibbons and S. W. Hawking, {\it Phys. Rev. D} {\bf 15} 2738 (1977).\\
$[7]$ B. Wang, Y. G. Gong and E. Abdalla, {\it Phys. Rev. D} {\bf 74} 083520 (2006).\\
$[8]$ Y. Gong, B. Wang and A. Wang, {\it JCAP} {\bf 01} 024 (2007).\\
$[9]$ R. S. Bousso, {\it Phys. Rev. D} {\bf 71} 064024 (2005); R.
G. Cai, H. S. Zhang and A. Wang, {\it Commun Theor. Phys.} {\bf
44} 948 (2005); M. Akbar and R. G. Cai, {\it Phys. Lett. B} {\bf
635} 7 (2006); {\it Phys. Rev. D} {\bf 75} 084003,
(2007).\\
$[10]$ M. R. Setare, {\it Phys. Lett. B} {\bf 641} 130 (2006); K.
Karami and A. Abdolmaleki, arXiv:0909.2427 [gr-qc]; H. M. Sadjadi,
{\it Phys. Rev. D} {\bf 73} 063525 (2006); N. Mazumder and S.
Chakraborty, {\it Class. Quant. Gravity} {\bf 26} 195016
(2009).\\
$[11]$ S. -F. Wu, B. Wang, G. -H. Yang and P. -M. Zhang, {\it
Class. Quant. Grav.} {\bf 25} 235018 (2008); A. Sheykhi and B.
Wang, {\it Phys. Lett. B} {\bf 678} 434 (2009); R. -G. Cai and N.
Ohta, {\it Phys. Rev. D} {\bf 81} 084061 (2010); N. Mazumder and
S. Chakraborty, arXiv: 1003.1606[gr-qc]; M. Jamil, A. Sheykhi and
M. U. Farooq, arXiv:1003.2093[hep-th]; A. Wang and Y. Wu, {\it
JCAP} {\bf 0907} 012 (2009); Q. -J. Cao, Y. -X. Chen and K. -N.
Shao, arXiv:1001:2597[hep-th]; R. -G. Cai, L. -M. Cao, Y. -P. Hu
and S. P. Kim, {\it Phys. Rev. D} {\bf 78} 124012 (2008); D.
Mateos, R. C. Myers and R. M. Thomson, {\it JHEP} {\bf 05} 067
(2007); R. -G. Cai, {\it Prog. Theor. Phys.} {\bf 172} 100 (2008);
R. G. Cai and L. -M. Cao, {\it Nucl. Phys. B} {\bf 785} 135
(2007); H. M. Sadjadi, {\it Phys. Rev. D} {\bf 76} 104024 (2007);
Q. -J. Cao, Y. -X. Chen and K. -N. Shao, arXiv:1001.2597; S.
Bhattacharya and U. Debnath, arXiv:1006.2600[gr-qc]; arXiv:1006.2609[gr-qc].\\
$[12]$ P. Szekeres, {\it Commun. Math. Phys.} {\bf 41} 55 (1975).\\
$[13]$ D. A. Szafron, {\it J. Math. Phys.}
{\bf 18} 1673 (1977).\\
$[14]$ D. A. Szafron and J. Wainwright {\it J. Math. Phys.}
{\bf 18} 1668 (1977).\\
$[15]$ J. D. Barrow and J. Stein-Schabes, {\it Phys. Letts.} {\bf 103A} 315 (1984).\\
$[16]$ S. Chakraborty and U. Debnath, {\it Int. J. Mod. Phys. D} {\bf 13} 1085 (2004)).\\
$[17]$ U. Debnath, S. Nath and S. Chakraborty, {\it Gen. Rel. Grav.} {\bf 37} 215 (2005 ).\\
$[18]$ U. Debnath, S. Chakraborty and J. D. Barrow, {\it Gen. Rel.
Grav.} {\bf 36} 231 (2004); U. Debnath and S. Chakraborty, {\it
JCAP}~ {\bf 05} 001 (2004).\\
$[19]$ S. Chakraborty, S. Chakraborty and U. Debnath, {\it Int. J.
Mod. Phys. D} {\bf 14} 1707 (2005); {\it Int. J.
Mod. Phys. D} {\bf 16} 833 (2007); {\it Gravitation and Cosmology} {\bf 13} 211 (2007). \\
$[20]$ Y. Zhang, Z. Yi, T. -J. Zhang and W. Liu, {\it Phys. Rev. D} {\bf 77} 023502 (2008).\\
$[21]$ P. Szekeres, \textit{Phys. Rev. D} \textbf{12} 2941
(1975).\\
$[22]$ R. -G. Cai and L. -M. Cao, {\it Phys. Rev. D} {\bf 75}
064008 (2007).

\end{document}